\pdfoutput=1
\RequirePackage{ifpdf}

\documentclass[cits,11pt,a4paper]{article}
\usepackage{jinstpub}

\usepackage{graphicx}
\usepackage{natbib}

\title{Neutrinoless Double Beta Decay with $^{82}$SeF$_6$ and Direct Ion Imaging}

\author[a]{D.R.~Nygren,}
\author[a,1]{B.J.P.~Jones,%
\note{Corresponding author.}}
\author[a,b]{N.~L\'opez-March,}
\author[c]{Y.~Mei,}
\author[a]{F.~Psihas}
\author[b]{ and J.~Renner.}

\affiliation[a]{Department of Physics, University of Texas at Arlington,
Arlington, Texas 76019, USA}

\affiliation[b]{Instituto de F\'isica Corpuscular (IFIC), CSIC \& Universitat de Val\`encia, 
Calle Catedr\'atico Jos\'e Beltr\'an, 2, 46980 Paterna, Valencia, Spain}

\affiliation[c]{Nuclear Science Division, Lawrence Berkeley National Laboratory, Berkeley, California 94720, USA}

\emailAdd{ben.jones@uta.edu}

\abstract{We present a new neutrinoless double beta decay concept: the high pressure selenium hexafluoride gas time projection chamber.  A promising new detection technique is outlined which combines techniques pioneered in high pressure xenon gas, such as topological discrimination, with the high Q-value afforded by the double beta decay isotope $^{82}$Se. The lack of free electrons in SeF$_6$ mandates the use of an ion TPC. The microphysics of ion production and drift, which have many nuances, are explored.  Background estimates are presented, suggesting that such a detector may achieve background indices of better than 1 count per ton per year in the region of interest at the 100~kg scale, and still better at the ton-scale.}


\begin{document}
\maketitle

\section{Introduction}

The search for neutrinoless double beta decay ($0\nu\beta\beta$) continues to enjoy high priority worldwide due to the unique possibility to reveal the Majorana nature of the neutrino.  Recent reviews describe the current experimental and theoretical status \cite{Ostrovskiy:2016uyx,DellOro:2016tmg, GomezCadenas:2010gs,gomez2011search}.  Discovery of this decay process would establish that the neutrino is its own antiparticle, a unique property among fermions, and that lepton number is not conserved.  Observation in any candidate isotope would be a major discovery as it would indicate physics beyond the Standard Model.

The quality of evidence for a discovery should, accordingly, be very robust.  It can be reasonably argued that two primary criteria should be imposed on any claim of discovery.  First, the fact that the observation of $0\nu\beta\beta$ would likely be based on a low statistics measurement (one or a handful of events in ton-scale detector given present exclusion limits) requires it to be effectively background-free.  This implies no contamination in the region of interest from either the relatively copious two neutrino double beta decay ($2\nu\beta\beta$) or from background radioactivity. The only known approach to reject the two-neutrino mode to the level of 0.1 ct/[ton yr FWHM] is to achieve, at the Q-value, an energy resolution better than 2\%  FWHM  with Gaussian characteristics, or still better in the presence of non-Gaussian tails.   Extraneous $\gamma$-ray, $\alpha$-particle, nuclear decays, neutron-induced reactions, and other well understood backgrounds can be rejected by several techniques including energy resolution, topological imaging, ion density, or daughter tagging. This latter class of background drives experimental design and generally imposes severe radio-purity requirements on materials and shielding.  
At present, no proposed experiments have met this criterion. 

A second criteria is the absence of signal when the active mass is replaced with identical material, except for replacement of candidate isotope with non-candidate isotope (i.e., enriched vs. depleted).  This requires the active mass to be easily interchangeable.  This was not possible, for example, in the Heidelberg-Moscow experiment \cite{KlapdorKleingrothaus:2001ke}, the reported signal from which required a further 11 years for refutation by an independent experimental program \cite{Auger:2012ar}.

Additional practical constraints in the design and deployment of a ton-scale $0\nu\beta\beta$ experiment include the need for an affordable active mass as well as a feasible and efficient detector. Affordability at the ton-scale requires a natural abundance at the $\sim$10\% level to minimize cost, as well as feasible enrichment using known techniques to a level of $\sim$90\% in isotopic composition.  A high detector efficiency at the ton-scale also favors monolithic technologies with uniform detector response, to minimize wastage of isotope used only for shielding.

In this paper, we propose a new 
$0\nu\beta\beta$ technology based on enriched selenium hexafluoride (SeF$_6$) gas, operated as the active medium of an ionic TPC \cite{Chinowsky:2007zz}. Despite several technological challenges, which we discuss throughout this paper, the advantages associated with SeF$_6$ as an active medium may yield a promising new, large-scale $0\nu\beta\beta$ search technique.  Throughout this work we will draw comparisons to double beta decay technologies based on xenon TPCs in either liquid \cite{Albert:2017owj} or gas \cite{Martin-Albo:2015rhw} phases.  This is because these are the most similar existing detectors to the concept being presented and comparison provides context for the performance metrics presented.  We outline of some of the issues to be faced in realization of such an experiment, and estimate its performance. 

\section{SeF$_6$ as a Neutrinoless Double Beta Decay Technology}

\begin{figure}[t]
\begin{center}
\label{Lego}
\includegraphics[width=0.8\columnwidth]{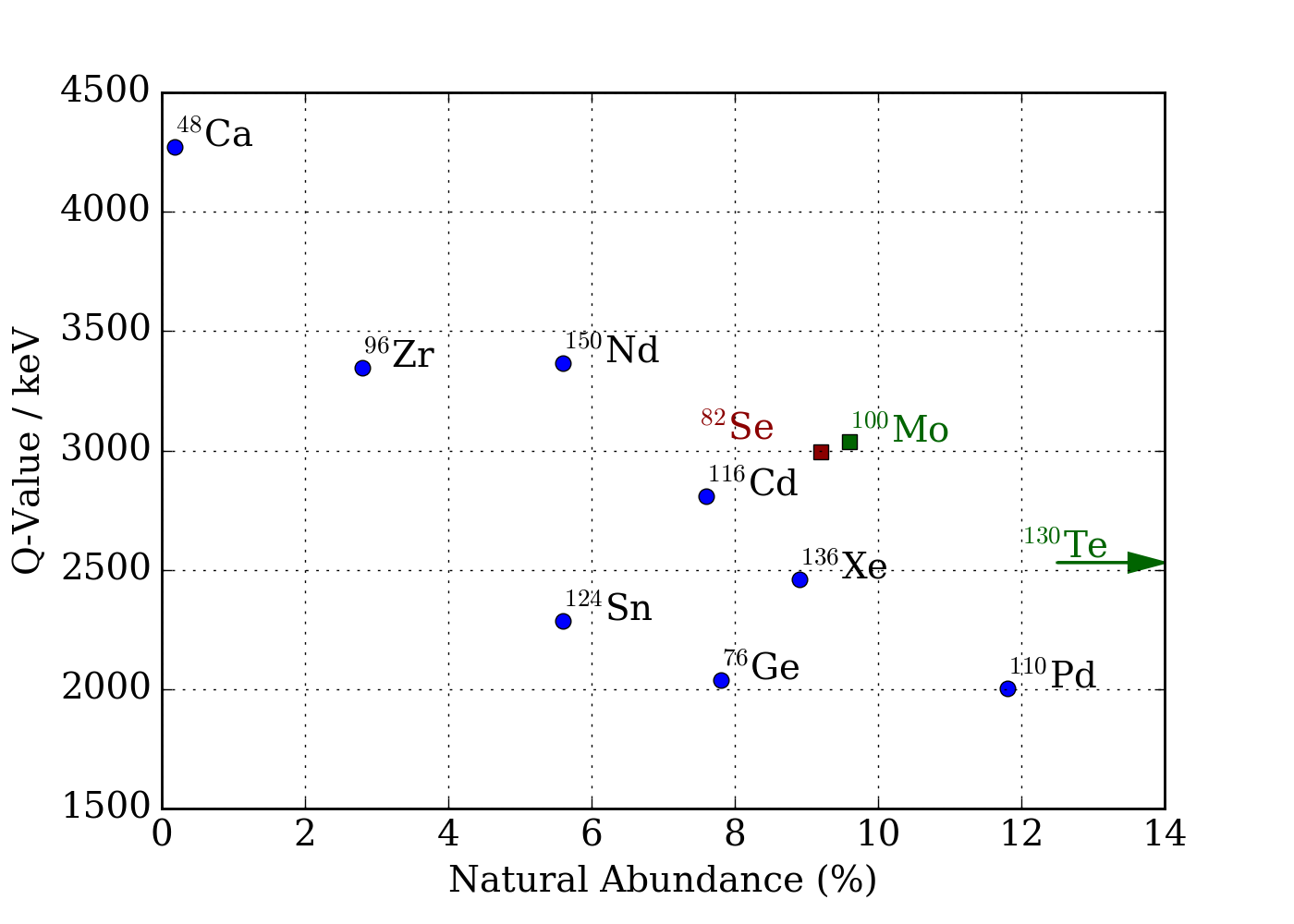} 

\caption{ Q-value vs. Abundance for various double beta decay isotopes. $^{82}$Se, the primary subject of this paper, is shown in red; the same detector could also be used with $^{100}$Mo and $^{108}$Te, shown in green. \label{fig:isotopes}}
\end{center}
\end{figure}

Several properties make $^{82}$Se attractive as a double beta decay isotope.  First, 
it has a relatively high Q-value.  At 2995 keV, the events-of-interest lie above the majority of the backgrounds from the $\gamma$ lines in the uranium and thorium chains, in particular falling well above the specific $^{214}$Bi and $^{208}$Tl lines which present serious challenges for xenon-based TPC experiments \cite{Martin-Albo:2015rhw,Albert:2017owj}. This represents a major advantage with respect to lower Q-value isotopes.  $^{82}$Se also has a high natural abundance of 9.2\%, which is comparable with both $^{136}$Xe and $^{76}$Ge.  The Q-values versus abundances of various double beta decay isotopes are shown in Fig.~\ref{fig:isotopes}, with $^{82}$Se highlighted.

It is noteworthy that the chemical form of SeF$_6$ is particularly convenient from the point of view of isotopic enrichment, since enrichment of $^{82}$Se typically begins with conversion to SeF$_6$ for gas-phase separation \cite{Beeman2015}.  The active background rejection afforded by gas phase TPCs ensures that the isotope is used efficiently, without large additional quantities of enriched isotope used for shielding (a benefit shared by both xenon and SeF$_6$ gas TPCs). Furthermore, because the active medium is a fluid, exchange between depleted and enriched isotope for confirmation of a signal would be a straightforward process, unlike in solid state technologies \cite{Agostini:2013mzu,Arnaboldi:2002du,Abgrall:2013rze}.  

Finally,
a detector designed for operation with SeF$_6$ could in principle be run with the similar compounds TeF$_6$ and MoF$_6$ with only minor modifications. The latter isotope also offers advantages of a high Q-value, of 3035 keV, and an electronically stable ionic daughter $^{100}$Ru$^+$, which may be a candidate for gas-phase daughter tagging \cite{Jones:2016qiq,McDonald:2017izm}.  These topics are outside the scope of the present work, but of interest for future studies.

\begin{figure}[t]
\begin{center}
\label{Lego}
\includegraphics[width=0.55\columnwidth]{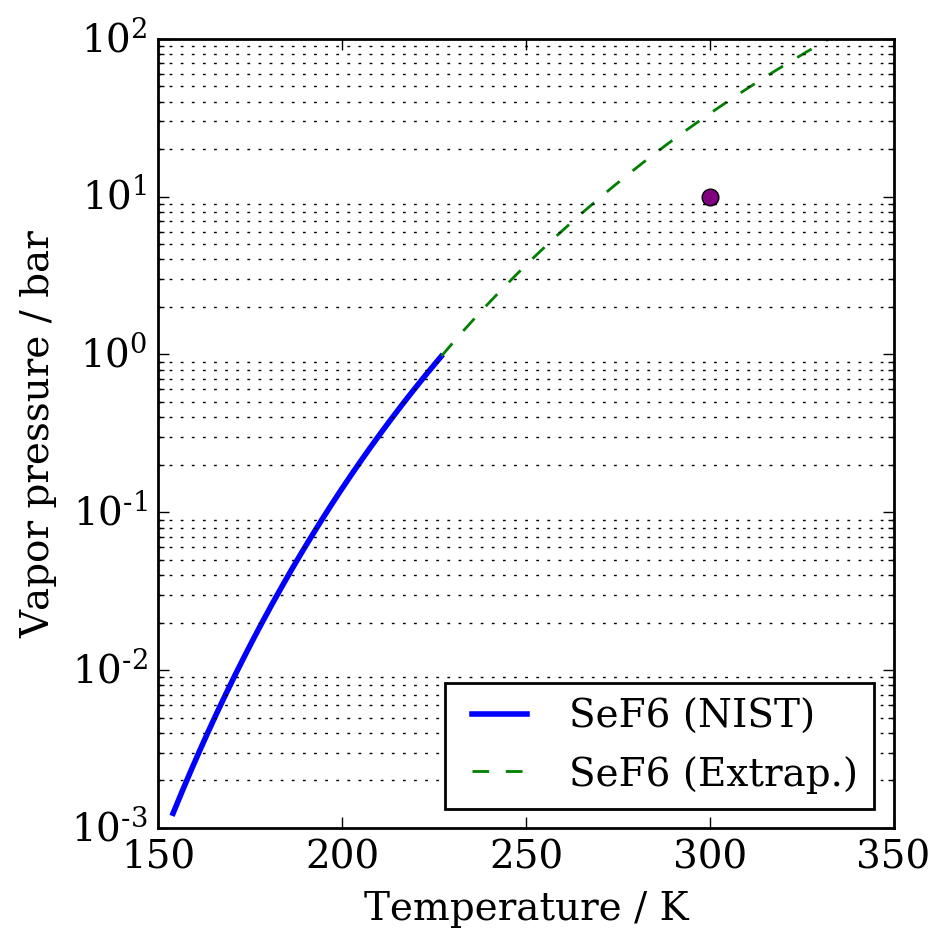} 

\caption{ Vapor pressure vs temperature for SeF$_6$, following the Antoine equation, using parameters reported at \cite{SeF6NIST}. The purple dot marks our assumed operating point, 10 bar at room temperature, which should be in the vapor phase. \label{fig:antoine}}
\end{center}
\end{figure}

The detector envisaged here is a high pressure gas TPC with topological imaging capabilities. It thus has much in common with the NEXT detectors \cite{Martin-Albo:2015rhw}, a program of high pressure xenon gas experiments geared towards mounting a low-background, ton-scale $0\nu\beta\beta$ search in $^{136}$Xe.  There are doubtless many specialized optimizations which can be made for an SeF$_6$ experiment. Here we explore a baseline set of operating parameters that appear reasonable, based on experience with xenon gas.    The Antoine curve for SeF$_6$, which shows the vapor pressure vs. temperature, is plotted in Fig. \ref{fig:antoine} using parameters from \cite{SeF6NIST}. This curve suggests that SeF$_6$ at room temperature will remain in the gas phase up to pressures around 30 bar, though notably, this involves some extrapolation from the measured range (154.5 - 227.4).  We chose 10 bar as a working pressure, since events in a 10 bar SeF$_6$ detector are expected to exhibit a similar topological structure to those in xenon at the same pressure, with some slight differences due to the higher Q-value and the different atomic composition of the gas.   A gaseous detector containing 1 ton of SeF$_6$ at 10 bar would have a density of 78 g/L and occupy 12 m$^{3}$ (around 2.3 m per side).

The two primary challenges of building a SeF$_6$ TPC are: 1) Toxicity. SeF$_6$ is highly toxic if inhaled, and corrosive if exposed to water or vapour \cite{SeF6MSDS}.  This complicates the process of deploying a large SeF$_6$ detector underground, requiring special safety controls; 2) Electron attachment. The high attachment coefficient of SeF$_6$ \cite{Jarvis} prohibits traditional electron-sensing TPC detectors, and instead mandates an ion TPC.  With emerging technologies such as Topmetal \cite{Topmetal} as described in Sec.~\ref{sec:Topmetal}, direct ion-imaging TPCs appear to be achievable.

Drift fields of 500-1000 V/cm appear reasonably applicable to the gas (which is a good HV insulator), to induce ion drift. Ions move comparatively slowly, however, which implies that larger fields may be beneficial to avoid long electronics integration times.

We imagine that both positive and negative ions can be in principle recorded using Topmetal, which is briefly reviewed in Sec. \ref{sec:Topmetal}.  Our primary focus in this work, however, is on the possibilities and limitations of SeF$_6$ as an active medium, rather than on the technological challenges of mounting an ion imaging TPC.  The latter include the usual aspects of radio-purity control, electronics integration and noise reduction.  The projections presented here suggest that further work in this direction is well motivated, although about this promising experimental concept remains to be explored.

\section{Microphyiscs of ion production and transport in SeF$_6$}

Only sparse data exists regarding the microphysics of SeF$_6$ gas \cite{Jarvis,Stockdale}.  However, some of the expected properties of an SeF$_6$ TPC detector can be inferred from a combination of the available literature and inferences drawn from the much better studied and structurally similar SF$_6$ molecule \cite{Phan,SF6Impact}.  The latter has been the focus of attention because of its industrial application as a non-toxic high-voltage insulator \cite{SF6Industry}. It has also been identified as a promising medium for an ionic TPC for dark matter searches \cite{Phan} and a component of the active medium for resistive plate chambers, added at sub-percent levels to inhibit streamer breakdown \cite{Capeans:2008gea,Capeans:2016eax,}.  Although quantitative information about SeF$_6$ inferred from SF$_6$ must be taken with caution, the qualitative properties relevant for the description of an SeF$_6$ TPC may be reasonably extrapolated.  Naturally, measurements to confirm these extrapolations are required, before the conclusions presented can be accepted with confidence.

We will use SeF$_6$ data where available, and augment this with information derived from SF$_6$ simulations in the MAGBOLTZ \cite{Magboltz} software package.  This package contains a full suite of cross sections for elastic and inelastic electron scatters and ionization processes in SF$_6$.  At the present time insufficient experimental data on SeF$_6$ cross sections exists to allow implementation as a medium in MAGBOLTZ.   As an example of typical TPC operating parameters we consider gas at a pressure of 10 bar with an applied electric field of 200-500 Vcm$^{-1}$ for double-beta-like events at the $^{82}$Se Q-value.   

The Fano factor for ionization is calculated to be F$_{\mathrm{SeF6}}\sim$ F$_{\mathrm{SF6}} =$0.19, approximately independent of voltage and pressure over parameter ranges 5-15 bar and  electric field strengths 100-1000 Vcm$^{-1}$.  
The degradation relative to the Fano factor of xenon, F$_{\mathrm{Xe}}\sim$ 0.16, stems primarily from the existence of various channels of ionization which have different ionization energies, as well as the low lying vibrational states of the molecular system.  
Because of the latter modes of energy loss, the W-value of SF$_6$ to produce electrons through direct ionization is higher than that of xenon, $W_{\mathrm{SeF6}}\sim W_{\mathrm{SF6}}=32$~eV, to be compared with $W_{\mathrm{Xe}}=22$~eV.  The intrinsic energy resolution that would in principle be achievable by collecting the direct ionization electrons is given by:
\begin{equation}
    \frac{\sigma_E}{E}=\sqrt{F \frac{W}{Q_{\beta\beta}}}\quad\quad\quad FWHM=2.355\sigma
\end{equation}

In SeF$_6$ gas the calculated resolution is 0.34\% FWHM at the Q-value for $^{82}$Se, to be compared with 0.28\% FWHM at the Q-value for $^{136}$Xe in xenon gas, with the higher Q largely compensating for the higher W and Fano factor.  These are both substantially better than the energy resolution achievable in liquid xenon, which uses partial cancellation of fluctuations between ionization and light \cite{Aprile:2007qd} to yield energy resolutions in the vicinity of 3\% FWHM \cite{Albert:2017owj}, depending on photon collection efficiency.

\begin{figure}[t]
\begin{center}
\label{Lego}
\includegraphics[width=0.99\columnwidth]{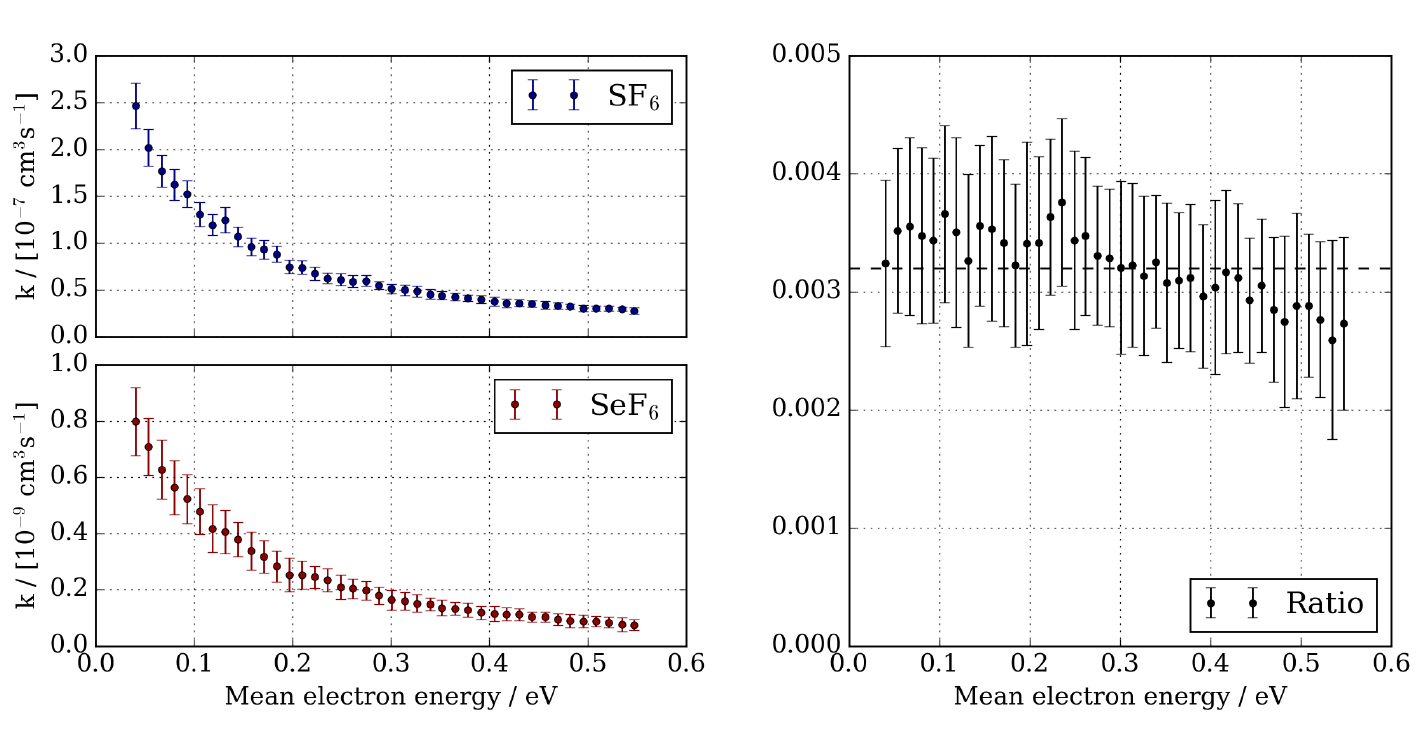} 

\caption{Left: Attachment coefficient as a function of energy for SeF6 and SeF6 in N$_2$ buffer gas from \cite{Jarvis}. Right: Attachment coefficient ratio as a function of energy. Within uncertainties of the measurement, the attachment ratio is consistent with the mean value of 0.0032. \label{fig:Attachment}}
\end{center}
\end{figure}

A property of SeF$_6$ that is well measured is its attachment cross section for low energy electrons.  Figure~\ref{fig:Attachment} shows data taken from \cite{Jarvis}, which presents the attachment coefficient as a function of mean electron energy for SeF${_6}$ and SF$_6$ in a nitrogen buffer gas.  Figure \ref{fig:Attachment},right, shows the ratio of these coefficients. Within the systematic uncertainty of the measurement, the ratio is consistent with the mean value of 0.0032 everywhere in the energy range. With this level of attachment, it is guaranteed that direct observation of free ionization electrons will not be possible.  Rather, these electrons will capture on neutral SeF$_6$ molecules to produce negative ion states near to the track core.  Because the attachment cross section peaks strongly at zero energy, it is expected that electrons will tend to thermalize before they capture. The negative ions thus produced will then drift slowly towards the anode where they can be collected.  Because the electrons thermalize before capture, the initial distribution of negative ions around the positive track core is likely to be similar to that of electrons around the track core in a typical high pressure gas TPC ionization even after thermalization.  At 10 bar, this is a region of around one micron in radius \cite{Nakajima:2015dva}.

Reduced mobilities of ions in SF$_6$ are in the range of  0.4-0.6 cm$^2$V$^{-1}$s$^{-1}$ \cite{Phan}, and we shall assume similar mobilities in SeF$_6$. This implies a drift time across a ton-scale experiment of order 0.5-1.0~s.  It also allows us to estimate the effects of ion-ion recombination near the track core - a vital consideration, given that substantial recombination, if present, could in principle spoil the energy resolution of such an experiment \cite{1997NIMPA.396..360B}. 

We now estimate, in an approximate manner, the expected extent of ion-ion recombination as the positive and negative ions separate. The assumed mobilities suggest that the time taken for positive and negative charge separation will be on the order of $\tau\sim0.5~\mu$s. Due to their large mass, the ions move with essentially thermal velocities $v=\sqrt{2 k_B T/m}\sim$ 150~ms$^{-1}$ during this time, and so cover a total distance of $d= v \tau \sim 75~\mu$m through random, thermal scattering, as they travel $~1\mu$m in the drift direction.  The expected number of interactions with an ion of the opposite charge within such a random walk is then  $\langle n \rangle = \sigma \rho d$. Given a 10~cm long, 1~$\mu$m radius track of energy 3~MeV and W=33~eV, the local ion number density inside the track core is $\rho\sim 3\times10^{12}$~cm$^{-3}$.  We assume that all ions are singly charged and so the cross section for capture is determined by the thermal Onsager radius, $R_{th}=e^2/4\pi \epsilon k_B T$ (58~nm) which gives a total geometrical capture cross section of $1.1\times10^{-10}$~cm$^2$  Thus the expected number of recombining collisions per ion is 0.023.  This implies that a mean of 2,100 recombination events will occur, with a binomial fluctuation of around 46.

This represents a total fluctuation on the collected ion yield of 0.05\%.  Such a treatment neglects the effects of local density fluctuations in the ionization density, which are known to enhance recombination fluctuations in liquid phase \cite{Dahl:2009nta}, but which we do not expect to be large in gas.  Comparison to the intrinsic fluctuations in ionization from the Fano factor, $\sim$140, suggests that recombination will be a sub-dominant but not entirely negligible source of energy fluctuations. 

Emerging from the interaction will be an ensemble of positive ions and an equal number of negative ions, each encoding equivalent information about energy and topology.  The exact identities of those ions is not yet known.  For the positive ionic species, we can infer the distribution based on available information for the SF$_6$ \cite{SF6Impact} and TeF$_6$ \cite{TeF6} systems, whose positive ion population distributions have been studied, though not in equivalent environments to our case of interest.

\begin{figure}[t]
\begin{center}
\includegraphics[width=0.99\columnwidth]{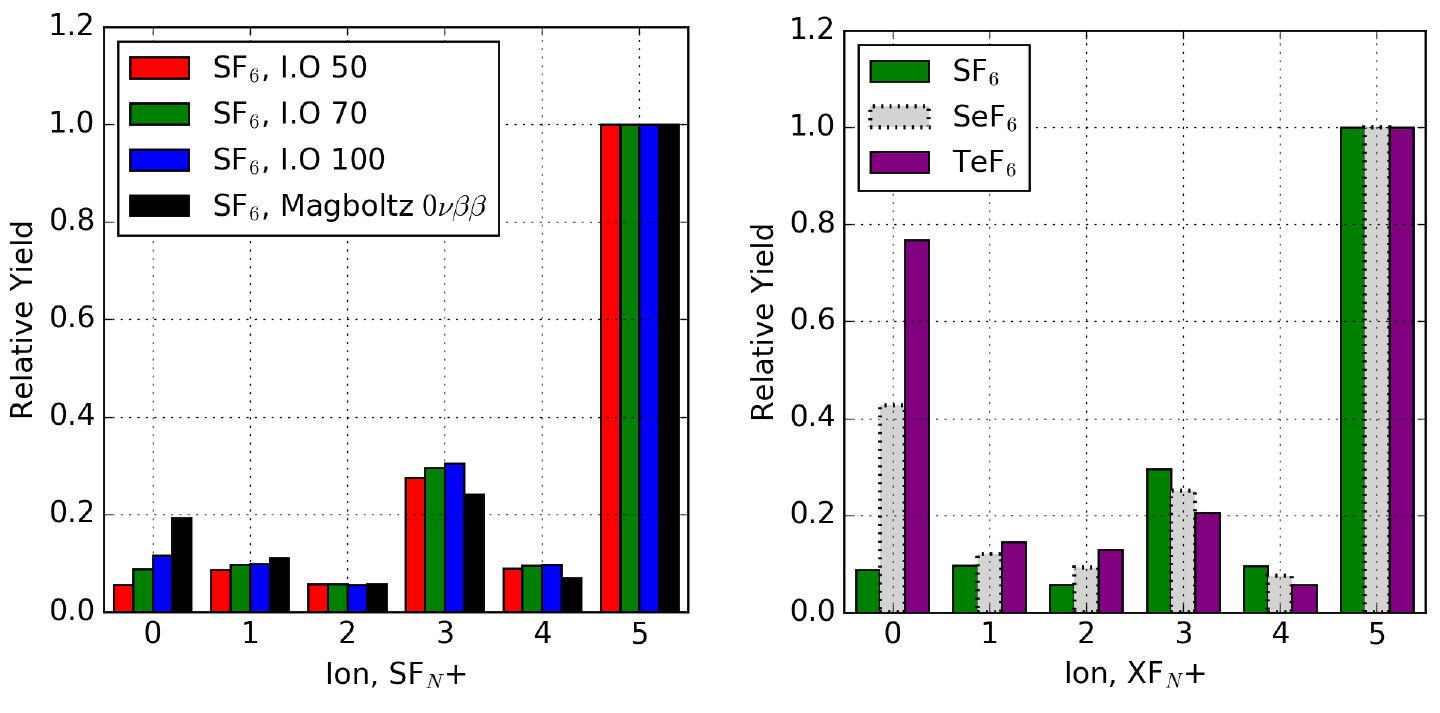} 

\caption{ Left: Population distribution of positive ion species made in ionization of SF$_6$, predicted by MAGBOLTZ in SeF$_6$ for double-beta-like events at the $^{82}$Se Q-value, and measured in electron impact ionization \cite{SF6Impact}; Right: Comparison of the electron impact ionization positive ion spectra for SF$_6$ (at 70V) and TeF$_6$ (from \cite{TeF6}), showing similar structure. As an estimate for the population of SeF$_6$ we take the midpoint between these two cases.  \label{fig:PopPlots}}
\end{center}
\end{figure}

For SF$_6$, we can also calculate the energy partition into different ionization channels to establish the positive ion populations that are produced in double-beta-like interactions using MAGBOLTZ.  We find contributions from all molecular ions in the series [SF$_N$]$^+$, with 0<N<5, with the dominant contribution from SF$_5^+$ and sub-leading contributions from other species.  Fragmentation spectra of SF$_6$ from impact ionization studies with 50, 70 and 100 eV electron beams reported in \cite{SF6Impact} show an almost identical structure, suggesting that the distribution over positive ions is approximately universal, and not strongly dependent on gas density and other environmental factors. These data are shown in Fig. \ref{fig:PopPlots}, left, each normalized to the intensity of the SF$_5$ peak to illustrate the shape of the distribution.  Electron impact ionization spectra from TeF$_6$ \cite{TeF6} show a similar trend, with the TeF$_5^+$ peak dominant and sub-leading contributions from all other charge states (Fig. \ref{fig:PopPlots}, right). The only substantial difference is the drastic increase in the Te$^{+}$ population over the S$^{+}$ population, which can be attributed to the lower ionization potential of Te (9.0~eV) relative to S (10.3~eV).  We can guess that SeF$_6$ will exhibit behaviour somewhere in between these two cases, lying vertically between them in the periodic table, and with an intermediate ionization energy for Se (9.8~eV).  As an estimate we pick a distribution half-way between the two, shown in Fig. \ref{fig:PopPlots}, right.

\begin{figure}[t]
\begin{center}
\includegraphics[width=0.99\columnwidth]{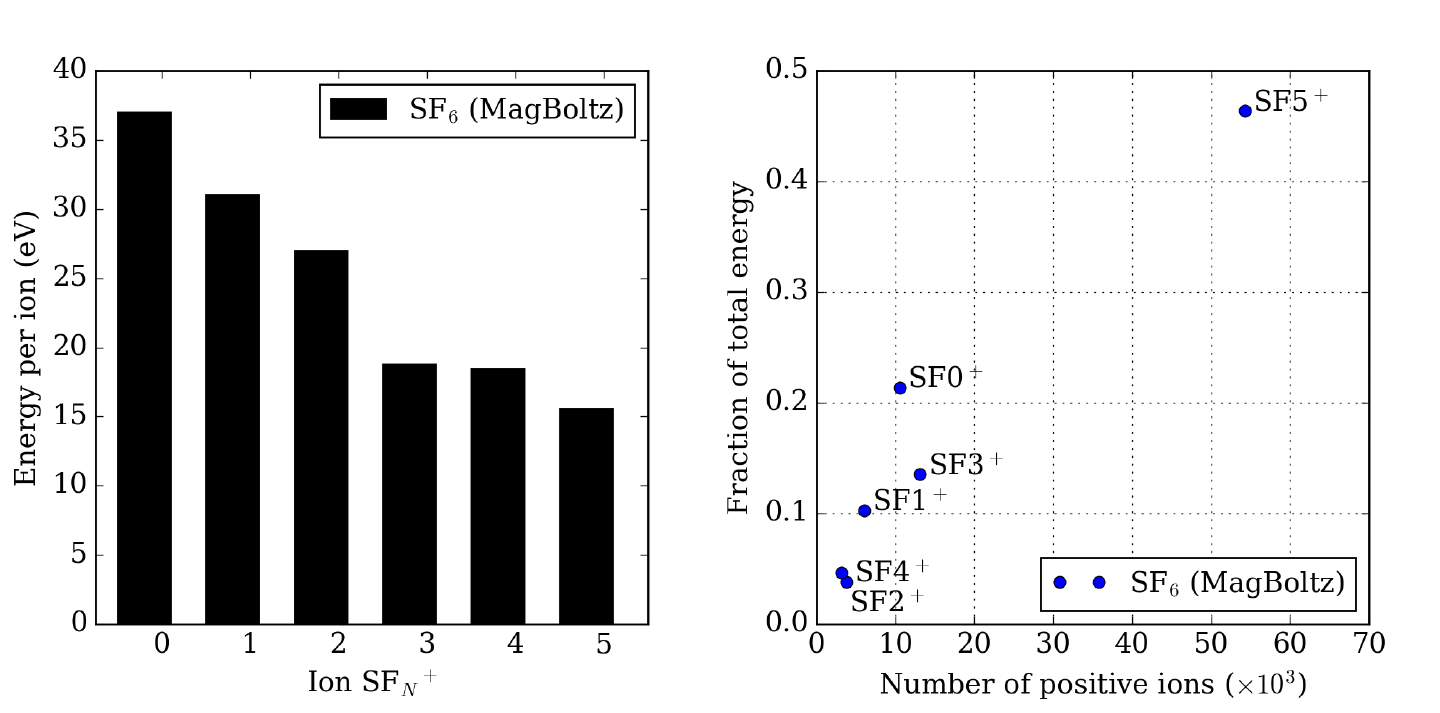} 

\caption{ Left: Effective energy cost per positive ion species in SF$_6$; Right: Fraction of observable energy vs total ion yield in SF$_6$. \label{fig:SF6PositiveDists}}
\end{center}
\end{figure}

The production mechanisms for negative ions are known to have somewhat different characteristics between SF$_6$ and SeF$_6$ \cite{Stockdale}.  In SF$_6$ there is a large cross section for production of a long-lived but meta-stable state SF$_6^{-*}$, which can either be collisionally stabilized to SF$_6^-$, or auto-detach into SF$_5^-$.  This leads to a distribution of anionic states between SF$_5^-$ and SF$_6^-$ which depends on pressure and drift field \cite{Phan}.  On the other hand, SeF$_6$ is not observed to form a similarly long-lived SeF$_6^{-*}$ ion \cite{Stockdale} in low pressure electron beam experiments. Thus, at relatively low pressures, the direct ionization reaction SeF$_6$+e$\rightarrow$SeF$_5^-$+F has been found to be dominant, yielding a population of only SeF$_5^-$ ions.  On the other hand, the swarm-based experiment at higher pressure conducted in nitrogen buffer gas \cite{Jarvis} find a population of 80:20 SeF$_6$:SeF$_5$ anions, suggesting that a stabilization mechansim for SeF$_6$ ions is present at higher pressure, though the authors of \cite{Jarvis} argue against a similar mechanism as observed in SF$_6$ by reference to selected ion flow tube measurements, and instead attribute the stabilization to interactions with the buffer gas.  Thus is it not entirely clear what balance of SeF$_6^-$ and SeF$_5^-$ ions should be expected in a 10 bar, high pressure pure SeF$_6$ gas.  It does, however, appear relatively likely that these will be the only two negative ion species present.

We also do not know the mobilities of SeF$_N^+$ or SeF$_N^-$ in SeF$_6$.  However, we can estimate how the mobilities of these ions will be related to each-other.  The ions are thermal, and as such, their instantaneous velocity is given by $v=\sqrt{2E/m}$, with E being approximately the thermal energy, kT.  As such, we might expect the mobility to scale like 1/$\mu\propto\sqrt{m}$.  In SF$_6$, where data is available, this naive model works relatively well. This scaling would suggest that the mobility difference between SF$_5^-$ and SF$_6^-$ would be around 7.2\%, whereas data show an $\sim$8\% difference \cite{Phan}.  Similarly, the naive scaling would suggest a mobility difference between SF$_5^+$ and SF$_3^+$ of around 19.5\%. Data \cite{SF3And5} show a 23.6\% difference.

\begin{figure}[t]
\begin{center}
\includegraphics[width=0.99\columnwidth]{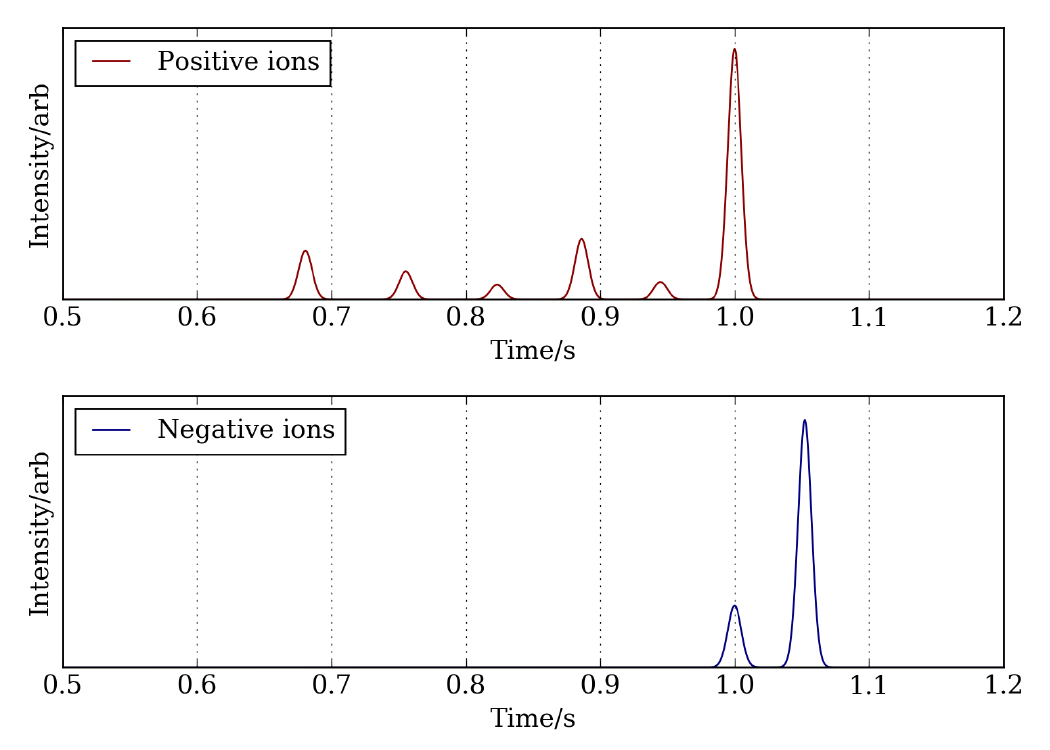} 

\caption{ Example pulse shapes that may be expected in the positive and negative ion channels after 1m drift with shaping time 5ms. \label{fig:ExamplePlusMinus}}
\end{center}
\end{figure}

Using this naive model, the mobilities of SeF$_N^\pm$ ions are likely to be much more similar to one another than SF$_N^\pm$ due to the heavier central atom.  However, with suitable electronics, assuming these mobilities are unique and charge-transfer reactions are not active, they may be resolvable.  Figure \ref{fig:ExamplePlusMinus} shows some example waveforms in the positive and negative ion channels. For the negative ions we take the distribution from \cite{Jarvis}, and for the positive ions, we use the intermediate distribution between SF$_6$ and TeF$_6$ described above.  We assume similar mobilities to SeF$_6$ ions in SF$_6$, equivalent mobilities for positive and negative species, and a 5~ms shaping time.  Given these assumptions, Fig. \ref{fig:ExamplePlusMinus} shows example pulse shapes that may be detected in the positive and negative ion channels after 1~m of drift.

The existence of these distinct sub-species raises the possibility of improving the energy resolution beyond that implied by the conventionally defined Fano factor.  In the case of the positive ions, because each ionization channel has a different effective W-value, appropriate weighting between them when integrating the collected charge may yield energy resolutions better than are obtainable by simple counting of quanta.  Using SF$_6$ as an example, and expecting SeF$_6$ to show a similar general trend, Fig. \ref{fig:SF6PositiveDists}, left shows the effective energy cost to make each type of positive ion, and Fig. \ref{fig:SF6PositiveDists}, right shows the population of ions produced in an event at $Q_{\beta\beta}$ versus the fraction of event energy which is used in production of each species. 

The case with the negative ions, on the other hand, is simpler. Here, initial ionization electrons produced with the same energy-cost-per-electron capture, to produce one of several possible negative ion states at random. Thus, while the population of SeF$_5^-$ and SeF$_6^-$ will be directly anti-correlated, each is associated with the same effective energy loss, and so no extra energy resolution information can be acquired beyond what is accessible from their sum.  By charge conservation, the total number of positive charges collected at the anode should be equal to the total number of negative charges collected at the cathode, with identical fluctuations.  Independent measures at the anode and cathode may thus improve achievable resolution by up to $\sqrt{2}$, in cases where instrumental noise rather than intrinsic ionization fluctuations are the limiting factor.  In cases where the intrinsic fluctuations are limiting, no further energy information can be obtained by collecting both species.

To explore the possible impact of independently weighting each ionization channel on the energy resolution, we developed toy model simulations based on the known energy partition into each state given by MAGBOLTZ. 
In our model, the energy was partitioned by simulating each interaction probabilistically, according to the total yields shown in Fig.~\ref{fig:PopPlots}, but also including unobservable states (i.e. vibrational and dissociative processes). Each interaction was assigned the corresponding energy costs as shown in Fig.~\ref{fig:SF6PositiveDists} for the ionization  channels and the initial state energy was randomly deposited, over many trials.
A limiting factor to the capabilities of our toy model comes from the treatment  of Auger cascades and secondary X-rays in MAGBOLTZ, which currently do not account for the total number of interactions in all sub-processes. 
Our model yields an effective $w_i$ and $F_i$ for each channel $i$ and reproduces the total Fano factor to within $\Delta f$ of a few percent, with a small dependence on the selected Auger model approximation.

We developed a maximum likelihood estimator (MLE) for the energy in a test sample of events with 
Q-value of 100~keV, using the effective resolution for each of the visible positive ion channels. Despite several approaches to optimizing energy measurement, the best achievable resolution was very similar to that implied by the naive Fano factor (within 0.01\% in resolution at 100~keV).   We thus conclude that despite some information on event energy residing in the ion population distribution, significant improvements beyond measuring the charge sum are not expected, and the ultimate resolution is likely to be limited by the conventionally defined Fano factor.  In practice, the intrinsic fluctuations of $\sim$0.4\% are likely to be sub-dominant to instrumental effects and to electronics noise.

\section{Direct Ion Imaging\label{sec:Topmetal}}

Due to its high electron affinity, free electrons drifting in SeF$_6$ are virtually nonexistent.  All electric charges are in the form of either positive or negative ions.  Positive ions cannot undergo well-controlled charge multiplication.  Under the proposed operating conditions, namely $10$~bar at room temperature, stripping electrons from negative ions in a high electric-field region and then multiplying them by avalanche gain is not a viable option either, unlike what has been demonstrated successfully in low-pressure (low-density) SF$_6$ gas~\cite{Phan}.  Using electrodes to directly collect ions and electronics to record the signal, without the help of charge multiplication, is the only recognized plausible scheme for charge readout in this scenario.  The charge readout must have three-dimensional imaging capability, i.e.\ having two-dimensional segmentation and detailed timing information for the arrival of ions, in order to utilize fully the event topology information.


\begin{figure}[t]
  \begin{center}
    \includegraphics[width=0.8\columnwidth]{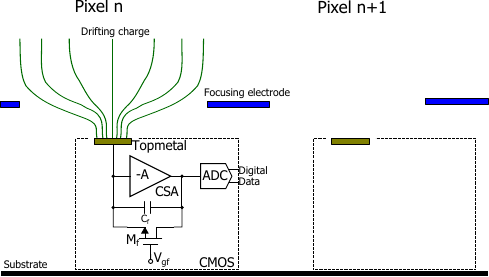}
    \caption{Sketch of Topmetal pixels that use CMOS integrated circuitry for direct charge collection, conversion and digitization, and an external electrode for charge focusing.  An exposed electrode in the topmost layer of CMOS (Topmetal, shown as a yellow rectangle) collects charge in gas directly.  Internal Charge Sensitive Amplifier (CSA) converts charge signal into voltage waveform, which is then digitized by the integrated ADC.  Digital signals are transmitted off the chip.  An elevated and separately biased electrode (blue rectangles) creates a focusing electrostatic field between itself and the Topmetal electrode (green curves).  A large plane can be filled with an array of these structures on a common substrate.  Not to scale.  \label{fig:Topmetal}}
  \end{center}
\end{figure}

We envision that an array of CMOS integrated charge sensors that directly collect charge and digitize the signal could satisfy all the requirements and enable an ion TPC with full 3D imaging capabilities.  Fig.~\ref{fig:Topmetal} shows the conceptual structure of a single pixel in the array.  The CMOS chip has an exposed electrode in its topmost metal layer (hence named Topmetal) that can directly collect charge from gas.  A low-noise charge sensitive amplifier (CSA) is placed right next to the electrode.  The CSA converts the charge into voltage signal which is then digitized by the internal ADC.  The following key factors support the effectiveness of this scheme:
\begin{description}
\item[Pixelation and charge focusing] The readout plane is fully pixelated in 2-dimensions.  A digital data network between CMOS sensors makes the engineering challenges manageable, therefore no wires or strips are used.  A focusing structure guarantees that  drifting charge will terminate on one of the Topmetal electrodes instead of getting lost in between pixels.  Such a structure could be as simple as a perforated metal sheet with the hole-pattern matching the desired pixelation.  The sheet shall be suspended above the sensor array and biased at an appropriate voltage.  The pitch between neighboring holes (pixels), $p$, determines the 2D resolution.  Besides minimal charge loss, the focusing structure allows the size (diameter) of the Topmetal electrode, $d$, to be significantly smaller than the pixel pitch $p$.  Since the capacitance of Topmetal-to-ground, $C_d$, which is positively correlated with the CSA noise, is proportional to $d^2$, it is important to minimize $d$.  The focusing structure allows this and we anticipate $d\lesssim1$mm and $p$ to be in the range from a few mm to about a cm.  The choice of $p$ is primarily constrained by diffusion and topological signatures.
\item[Low-noise CSA with long signal retention] Having low electronic noise at every pixel is essential to achieving necessary total energy resolution at $Q_{\beta\beta}$.  With $d\lesssim1$~mm, the capacitance as seen by the CSA, $C_d$, is on the order of a few pF and is completely internal to the CMOS process.  CSA design using a modern CMOS process to achieve an Equivalent Noise Charge (ENC) of $<50$~e$^-$ with such $C_d$ is readily available.  Since a double-beta event will spread charge to a few hundred pixels and the noise from pixels that see charge my be reasonably expected to add up quadratically, we expect that the electronics will contribute less than $10^3$~e$^-$ equivalent noise.  

A $<2\%$ FWHM energy resolution is readily achievable with such additional noise.  Detecting drifting ions poses additional requirements on CSA.  Since ions drift at a speed on the order of mm/ms, the CSA must retain signal greater than ms time scale and the noise performance must not deteriorate at ms integration time.  By using fF scale feedback capacitance, $C_f$, and a close-to-shutoff but well-controlled transistor, $M_f$, as the feedback resistor inside the CMOS sensor, it has been demonstrated in \cite{Topmetal} that direct observation of drifting ions in gas can be achieved with CMOS Topmetal sensors.  The equivalent resistance of $M_f$ is fine-tuned by an internally generated voltage $V_{gf}$.  The $RC$ constant that is determined by $C_f$ and the equivalent resistance of $M_f$ is tuned to well beyond $\sim$ms, which helps retain the slow-drifting ion signals on CSA.
\item[CMOS integration] The scale of the detector requires a meter-sized charge readout plane with $\sim10^5$ pixels on the plane.  The charge sensing scheme must be able to scale up to the required size both technologically and cost-effectively.  CMOS Integrated Circuit (IC) process is a technology of choice in this scenario.  CMOS allows the integration of all required circuitry, including the CSA, the ADC, and the digital control and signal processing, to be very close to the site of charge sensing: the Topmetal electrode.  The tight integration between the electrode and the CSA, as well as the direct on-site digitization minimizes the external noise pickup and effects such as microphonics that has plagued conventional wire readouts.  All sensors form a digital network to communicate digital data.  It reduces the otherwise unmanageably large number of signal wires to a handful of lines that eventually drive data through the chamber via feed-throughs.  The scheme described here utilizes entirely commercial CMOS process without any deviation from or addition to the industrial standards.  It enables the direct and complete leveraging of microelectronics industry for cost-effective scaling.  The consistency in performance of pixels and the uniformity of the assembled plane are guaranteed by the industrial quality control practices.  Regarding the material radiopurity, the substrate shall be built using Kapton Printed Circuit Board (PCB), which is a known clean material used by many low-background experiments.  Silicon based CMOS sensors are clean as well.  With proper design, the CMOS sensors can operate without any external components such as capacitors and resistors.  CMOS sensors are wire-bonded to the substrate, eliminating solder joints, which is a usual culprit for radioactive background.  Detector components other than the active top-metal elements are largely similar to those used in high pressure xenon gas TPCs, which are projected to achieve adequate ratio-purity for nearly background-free 100~kg-scale experiments \cite{Martin-Albo:2015rhw}.  High isotopic enrichment of selenium has been demonstrated via centrifugal separation of SeF$_6$ \cite{Beeman2015}, and natural fluorine comprises 100\% abundance of $^{19}$F with no long-lived radioisotopes, eliminating another possible source of radioimpurity.

\end{description}

It is worth emphasizing that signal waveforms shall be digitized by the internal ADCs.  A conventional pulse shapers is not used here.  Due to the slowness of ion drifting, a shaper with ms time constant would be required to recover the energy information.  It is difficult to design such an analog shaping in CMOS and it would be power-hungry.  Instead, the waveform of the CSA output shall be digitized directly.  The $z$ information shall be recovered by processing the rising edge of the CSA output waveform and the energy information shall be recovered by an equivalent digital shaper.  The ADC sampling time interval can be as long as sub-ms to satisfy relevant spatial and energy resolution requirements.

\section{Quality of Topological Signature in SeF$_6$}

\begin{figure}[b!]
\begin{center}
\includegraphics[width=0.99\columnwidth]{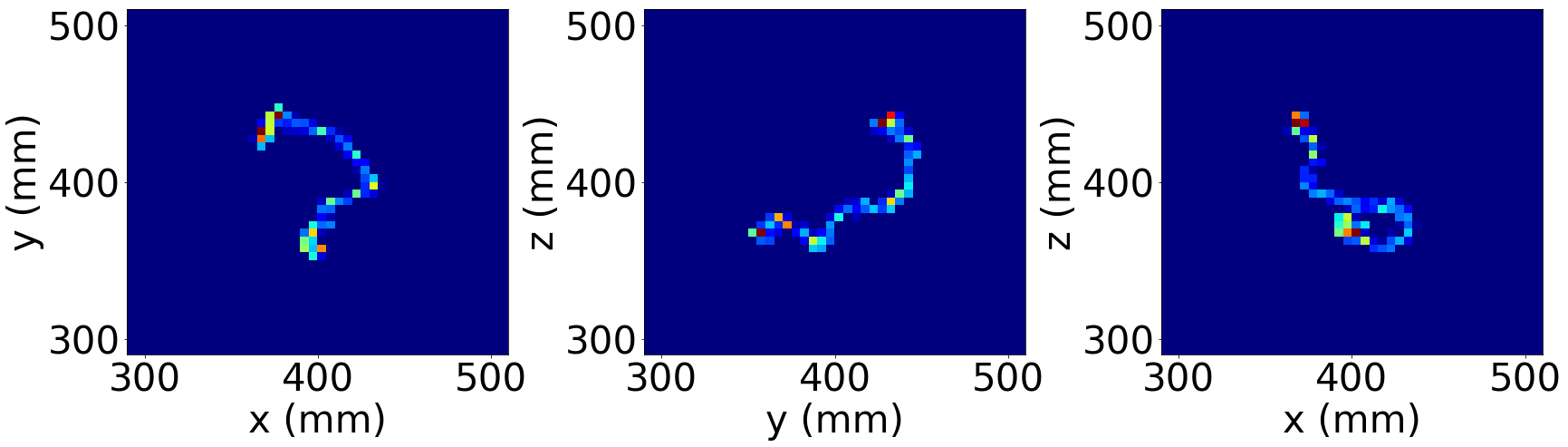}
\includegraphics[width=0.99\columnwidth]{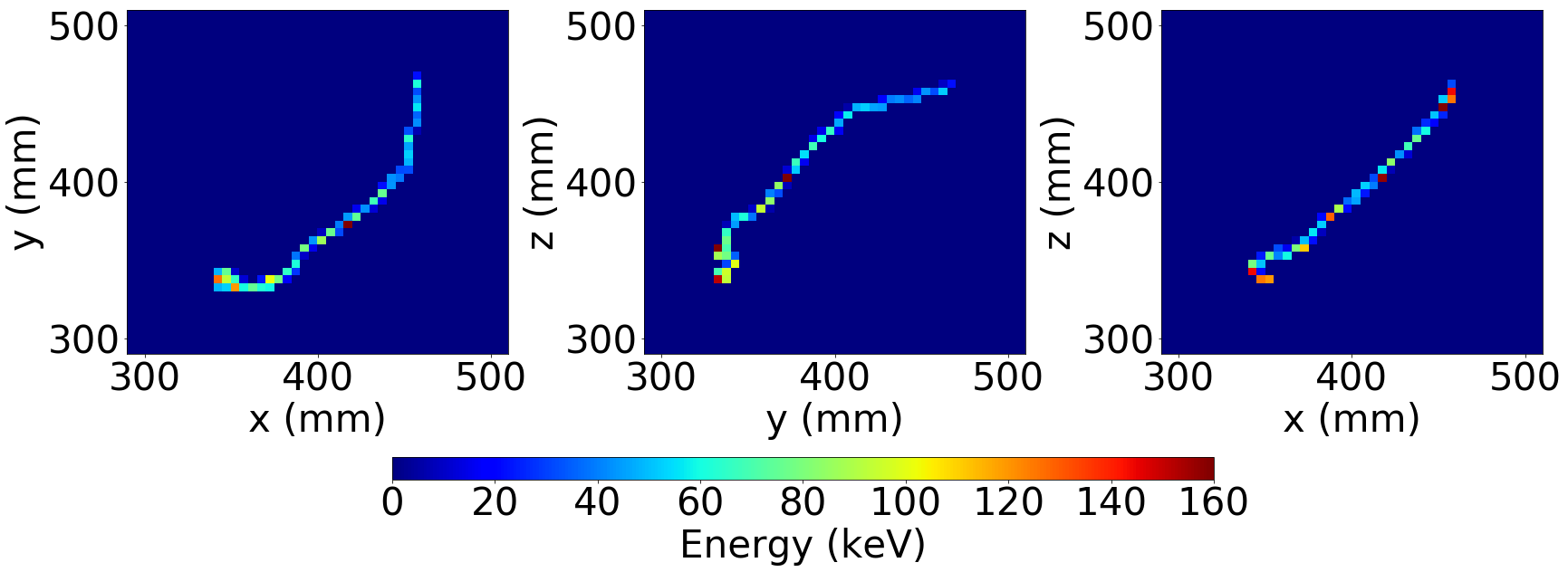}

\caption{An example of an SeF$_6$ double beta decay event (top), voxelized with 5~mm resolution and shown from three orthogonal viewing angles.  The two-blob signature is clearly visible.  A single-electron event (bottom) of the same energy is also shown. \label{fig:SeF6Event}}
\end{center}
\end{figure}

Reconstruction and classification of event topology is a powerful discriminant between the double beta decay signal and radioactive backgrounds in high pressure gas TPCs \cite{Ferrario:2015kta}.  The topological discrimination has effectively two components: 1) Single vs multi-site classification: most true signal events have all the energy localized in one contiguous track, whereas gamma rays and neutron backgrounds often produce multiple, independently localized deposits in the detector; 2) Single vs multi-electron discrimination, using the fact that one-electron events have a ``blob'' at one end where the electron slows down and repeatedly scatters, leading to a higher energy density, whereas two-electron events have two ``blobs'', indicating the double electron topology.  Given that ions experience far less diffusion than electrons in room temperature gas, ion TPCs can in principle access much higher quality topological information than traditional electronic TPCs. This reduction in diffusion constant $D$ is codified in the Einstein relation $D=\mu k_B T / q$, where $T$ is the effective temperature of the drifting species, $\mu$ is the mobility and $q$ is the charge  of the drifting species.  The average energy of electrons drifting in high pressure xenon gas is significantly larger than thermal (0.6 eV at 200 V/cm), whereas the ion energies remain thermalized ($\sim$0.02 eV at 300 K) due to their small relative acceleration between collisions.  The extent to which charges diffuse over a fixed distance is proportional to $\sqrt{k_B T}$, so ions are expected to experience diffusion reduced by a factor of at least 5.5 relative to electrons drifting over a similar distance, with higher suppression at larger drift fields.  Spatial resolutions of such detectors will be limited in practice by the pixel size of the readout, rather than by diffusion.

Given a similar pixel size, the topological discrimination capability achievable in SeF$_6$ is further expected to be somewhat improved beyond than that achievable in xenon for two reasons: 1)~the multiple scattering length of electrons in SeF$_6$ is slightly longer due to the lower Z of the constituent atoms, and 2) the Q-value is higher, producing longer tracks.  An example of a signal event simulated in SeF$_6$ using GEANT4 is shown in Fig. \ref{fig:SeF6Event}, top, with the two electron blobs clearly discernible.  A background event at the same energy is shown in Fig. \ref{fig:SeF6Event}, bottom, and does not exhibit this characteristic.

\begin{figure}[t]
\begin{center}
\includegraphics[width=0.6\columnwidth]{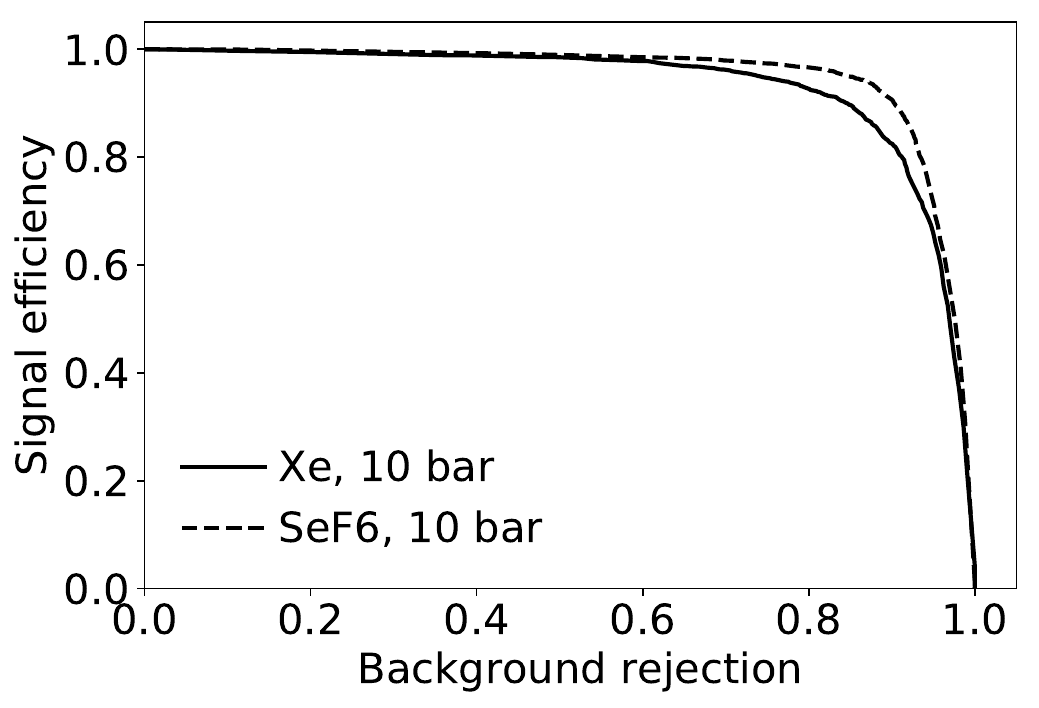}

\caption{Signal efficiency vs background rejection as a function of varying cut strength using a DNN-based topological analysis  \label{fig:DNNPlot}}
\end{center}
\end{figure}

Applying the techniques developed for topological reconstruction using Deep Neural Networks (DNNs) in \cite{Renner:2016trj}, we have compared the topological power achievable using the topological signature in Xe to that in SeF$_6$, assuming similar spatial resolution in the readout system.  Naturally, the background rejection capability is a function of the allowable signal inefficiency as less golden signal events are rejected. The signal efficiency as a function of background rejection for Xe and SeF$_6$ events, both reconstructed with position resolution of 3~mm is shown in Fig. \ref{fig:DNNPlot}.  As expected the achievable performance in SeF$_6$ is slightly better than in xenon.

Monte Carlo events were simulated using GEANT4 in a volume of pure gas at 10 bar.  $0\nu\beta\beta$ events were generated at $Q_{\beta\beta,Xe} = 2.46$ MeV for $^{136}$Xe and $Q_{\beta\beta,Se} = 2.99$ MeV for $^{82}$Se using the DECAY0 code \cite{Ponkratenko_2000} to generate theoretical distributions of the momenta of the two product electrons.  In each case, single-electron ``background'' events at the same energy were also generated. The ionization tracks of these events were saved as a series of hits (x,y,z,E) with a maximum step size between hits of 1 mm.  The energy and position of these hits were then smeared in a manner consistent with an energy resolution of 0.8\% FWHM at the respective Q-value and transverse and longitudinal diffusion of 3 mm/$\sqrt{\mathrm{cm}}$, assuming a maximum drift length of 500 mm.  The energy contained in the hits was then used to fill cubical spatial elements of size 5 mm $\times$ 5 mm $\times$ 5 mm, and the projections these elements in x-y, y-z, and x-z were then converted to an RGB image, in which each of the three projections was assigned to one color channel. 

The Nvidia DIGITS \cite{DIGITS} software interface to the Caffe \cite{jia2014caffe} deep learning library was then employed to train a deep neural network, the GoogLeNet \cite{Googlenet}, to classify events into two categories, ``signal'' and ``background,'' based solely on the RGB image of their projections. The dataset, consisting of 45000 signal events and 45000 background events, each for Xe and SeF$_6$, was divided in each case into training events (36000), validation events (4500) for monitoring the training process and test events (4500) for final verification of the accuracy of the network. The network returned a probability of classification as signal or background, and varying the threshold of event acceptance for the test events traces out the curve shown in Fig. \ref{fig:DNNPlot}.

Notably the effects of the differences in signal formation (multiple ion mobilities, ion-ion recombination, etc) and different instrumentation (Topmetal vs. electroluminescence) have not been included in the estimates of this section.  Properly accounting for these effects would require detailed microphysical and instrumental simulations, and at the present time, the inputs to these simulations are not well known. The key conclusion, however, is that the topological discrimination achievable in SeF$_6$ is slightly better than in high pressure xenon gas, if the ionization image can be equally effectively imaged through ion counting rather than electron counting.  Due to the reduced diffusion expected in the ion channel, this appears plausible.  On the other hand, long integration times and coexistence of multiple ion species may have detrimental effects.  In the latter case, deconvolution should be applicable to collapse the image, consisting of several ionic echoes arriving at distinct times, into a single image at each end of the detector. The performance of such a procedure is beyond the scope of the present work.

\section{Expected backgrounds in a SeF$_6$ gas-phase TPC \label{sec:Backgrounds}}

\begin{figure}[t]
\begin{center}
\includegraphics[width=0.99\columnwidth]{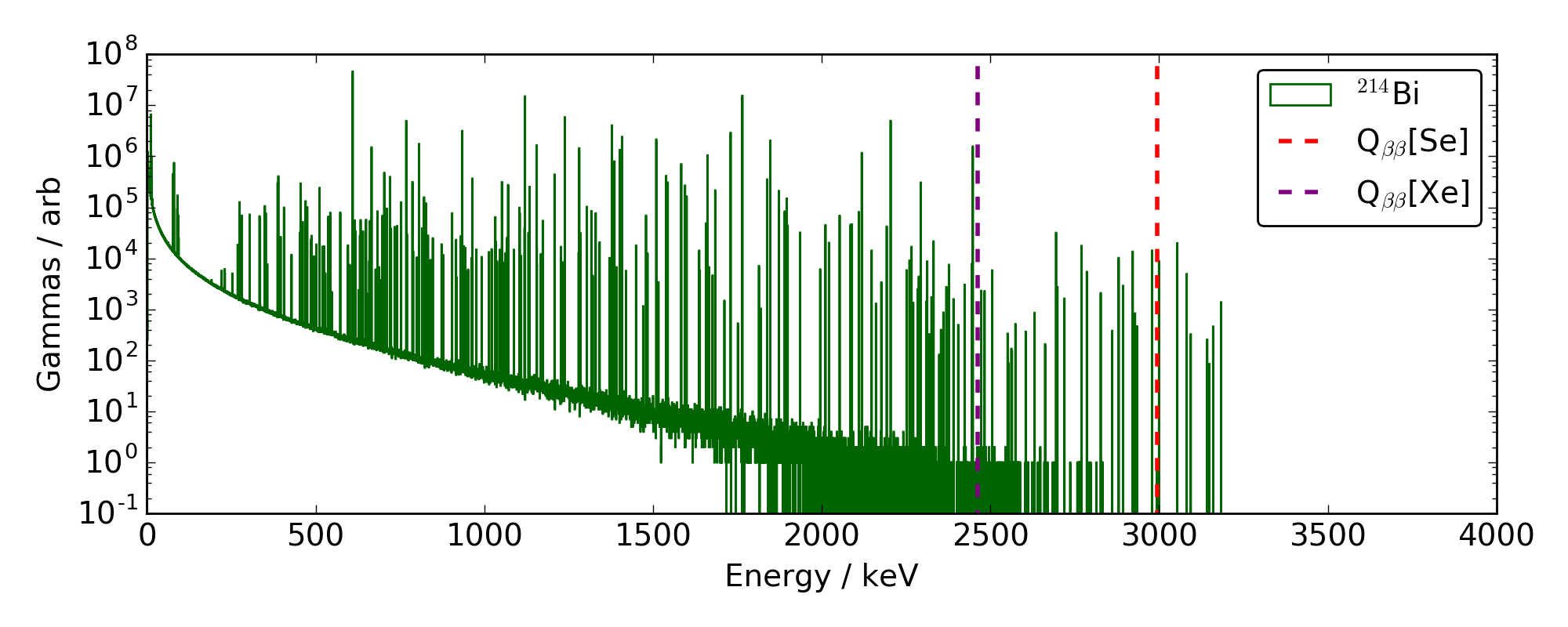}
\includegraphics[width=0.99\columnwidth]{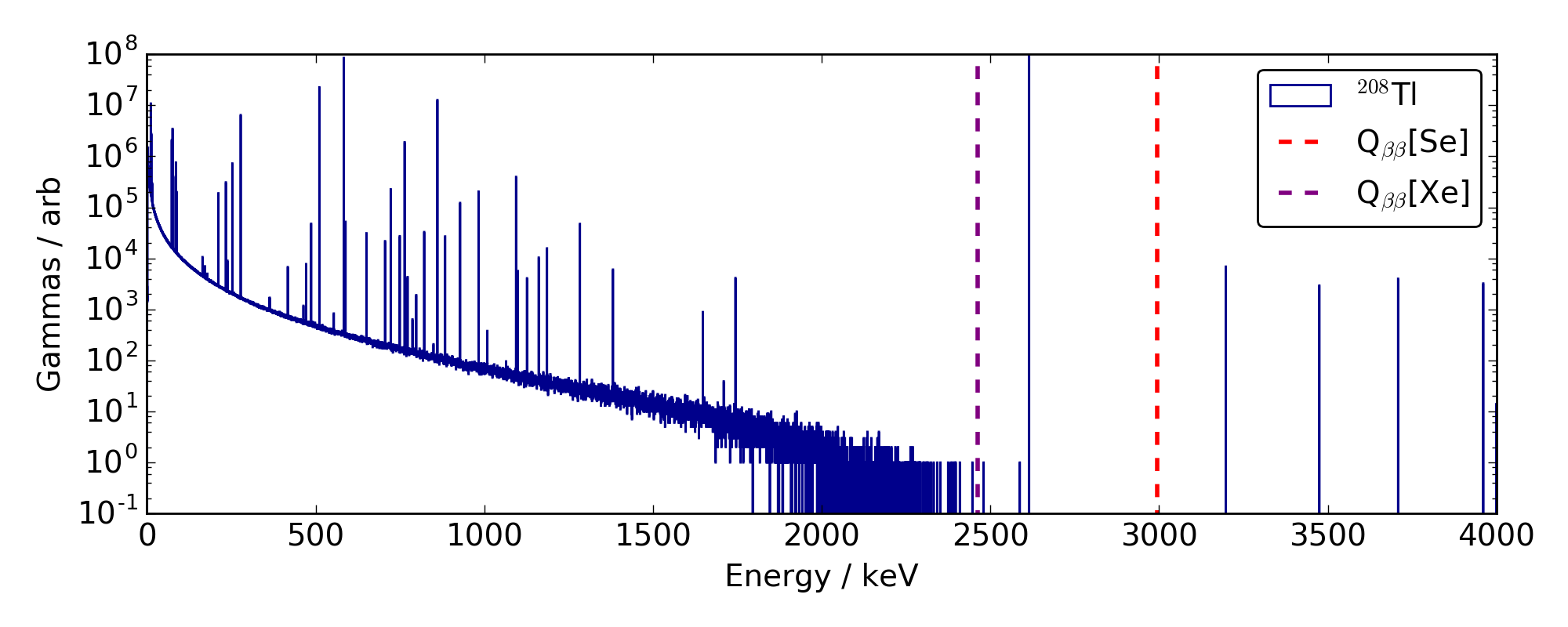}

\caption{ Spectra of gamma ray backgrounds from $^{214}$Bi and $^{208}$Tl decays with Q-values of $^{136}$Xe and $^{82}$Se marked. \label{fig:BackgroundSpectra}}
\end{center}
\end{figure}

The important backgrounds in a SeF$_6$ gas TPC differ from those in a HPGXe TPC in that they are in a different energy range; However, the observables are similar - event topology and energy; and the physical structure of the device will also be similar, employing a pressure vessel containing high pressure gas, a field cage, and at least one plane of silicon devices.  This means that the background models developed for high pressure gas TPCs may be plausibly extrapolated to SeF$_6$ to obtain order-of-magnitude level estimates for expected background rates. In this section we estimate the expected level of background in a SeF$_6$ TPC, making the assumption that the structure of the device is equivalent to the NEXT-100 high pressure xenon gas TPC.

In Ref.~\cite{Martin-Albo:2015rhw}, a detailed study of the expected background in NEXT-100.  The prediction was 7 counts per ton per year, per 0.5\% FWHM.  The dominant backgrounds are from $^{214}$Bi gamma rays, with a sub-leading contribution from $^{208}$Tl gamma rays.

\begin{figure}[t]
\begin{center}
\includegraphics[width=0.49\columnwidth]{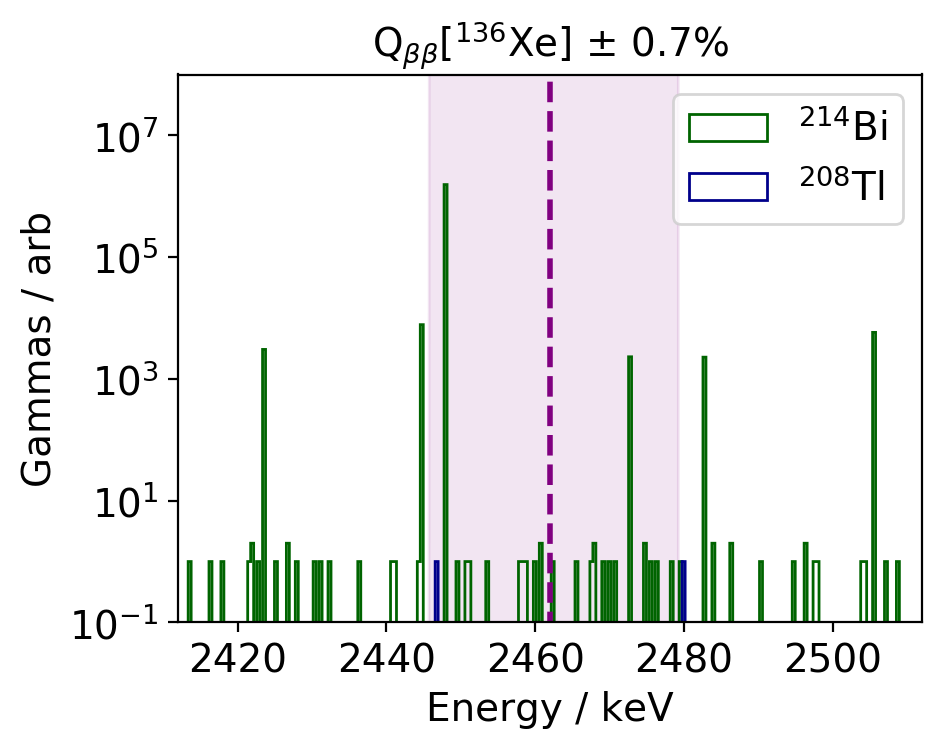}
\includegraphics[width=0.49\columnwidth]{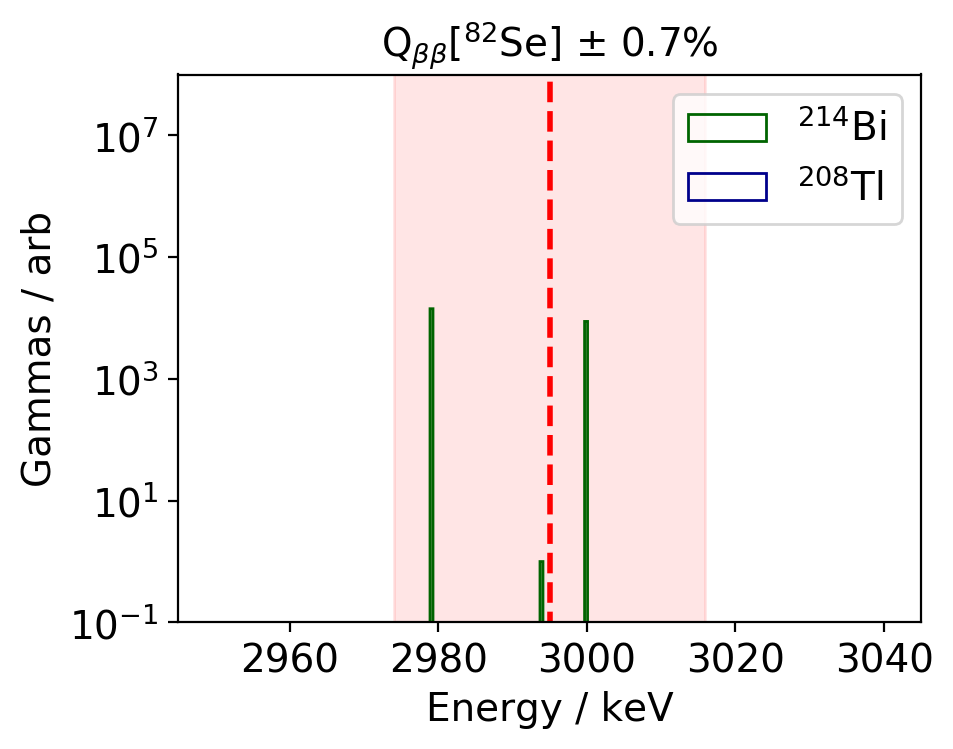}

\caption{Backgrounds in the vicinity of the Q-values of $^{136}$Xe and $^{82}$Se.  Regions corresponding to width 0.7\% FWHM around the Q-value, the demonstrated energy resolution of the running NEXT-NEW TPC, are marked.   \label{fig:BackgroundSpectraClose}}
\end{center}
\end{figure}

In SeF$_6$, different, higher energy gamma lines contribute to the backgrounds.  Inspection of the lines in the natural uranium and thorium chains shows that again, only  $^{214}$Bi and $^{208}$Tl are likely to contribute significantly. The energy spectra of gamma rays produced in the decays of these isotopes over the full energy range are shown in Fig.~\ref{fig:BackgroundSpectra}, with the Q-values of $^{136}$Xe and $^{82}$Se marked.  In both cases, the backgrounds are visibly much reduced in the energy range of interest for $^{82}$Se, relative to $^{136}$Xe.  This is shown more clearly in Fig.~\ref{fig:BackgroundSpectraClose}, which presents a sub-region around the Q-value. A colored band in each case shows an energy resolution of 0.7\% around the Q-value, which is the presently demonstrated extrapolated energy resolution from the NEXT-NEW experiment \cite{Ferrario:2017zqp} . Given similar energy resolution, no lines in the $^{208}$Tl spectrum are within the region of interest for $^{82}$Se. Furthermore, inspection of the wider energy range shows that down-scattering of higher energy gammas from $^{208}$Tl, which provide the major contribution to the $^{208}$Tl background in high pressure xenon gas, will be at least four orders of magnitude suppressed in $^{82}$Se relative to $^{136}$Xe due to the absence of strong lines at those energies.  Backgrounds lines from $^{214}$Bi are present for $^{82}$Se, but at much lower levels than the corresponding backgrounds in $^{136}$Xe. It is clear that the expected background rates in a SeF$_6$ gas TPC would be significantly lower than in high pressure xenon gas TPCs, which are already approximately two orders of magnitude lower than similarly sized liquid xenon experiments \cite{Albert:2017owj}.

\begin{figure}[t]
\begin{center}
\includegraphics[width=0.49\columnwidth]{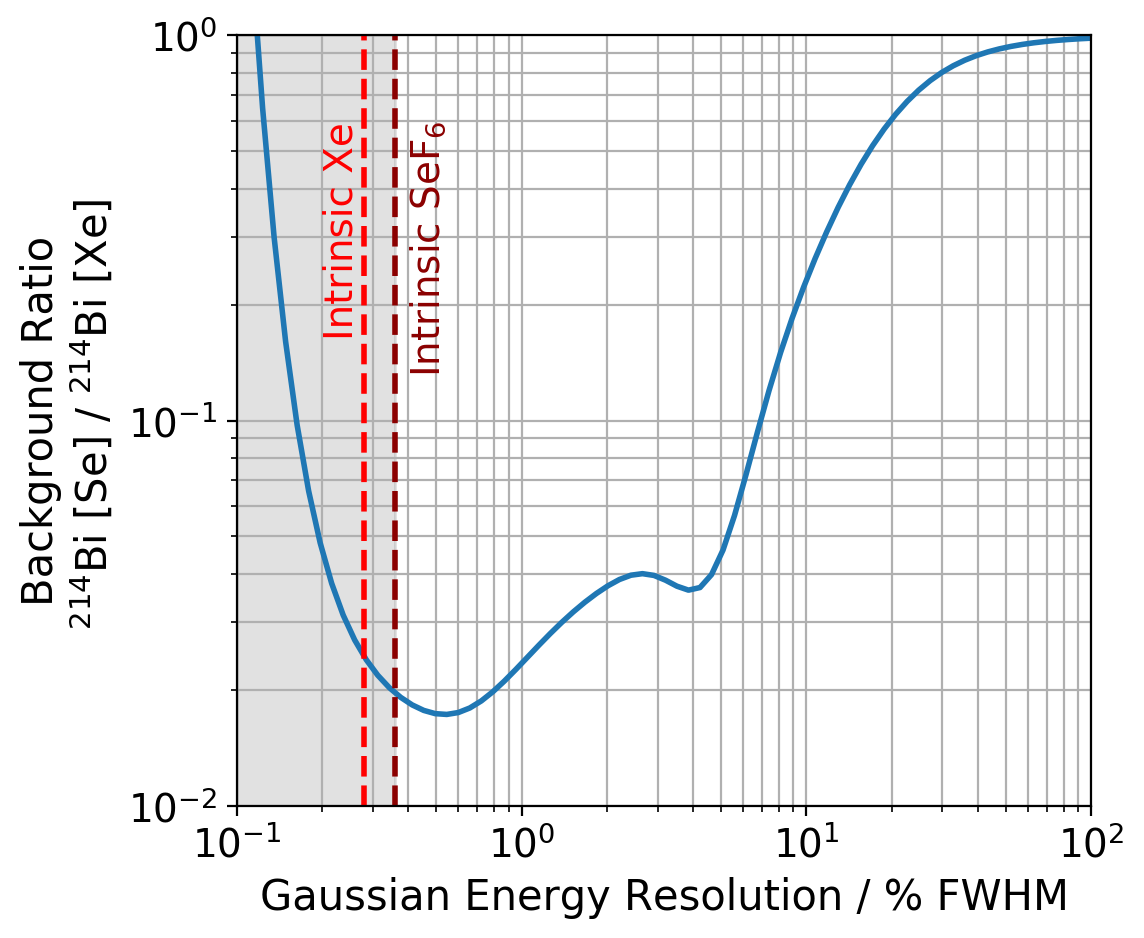} 
\includegraphics[width=0.49\columnwidth]{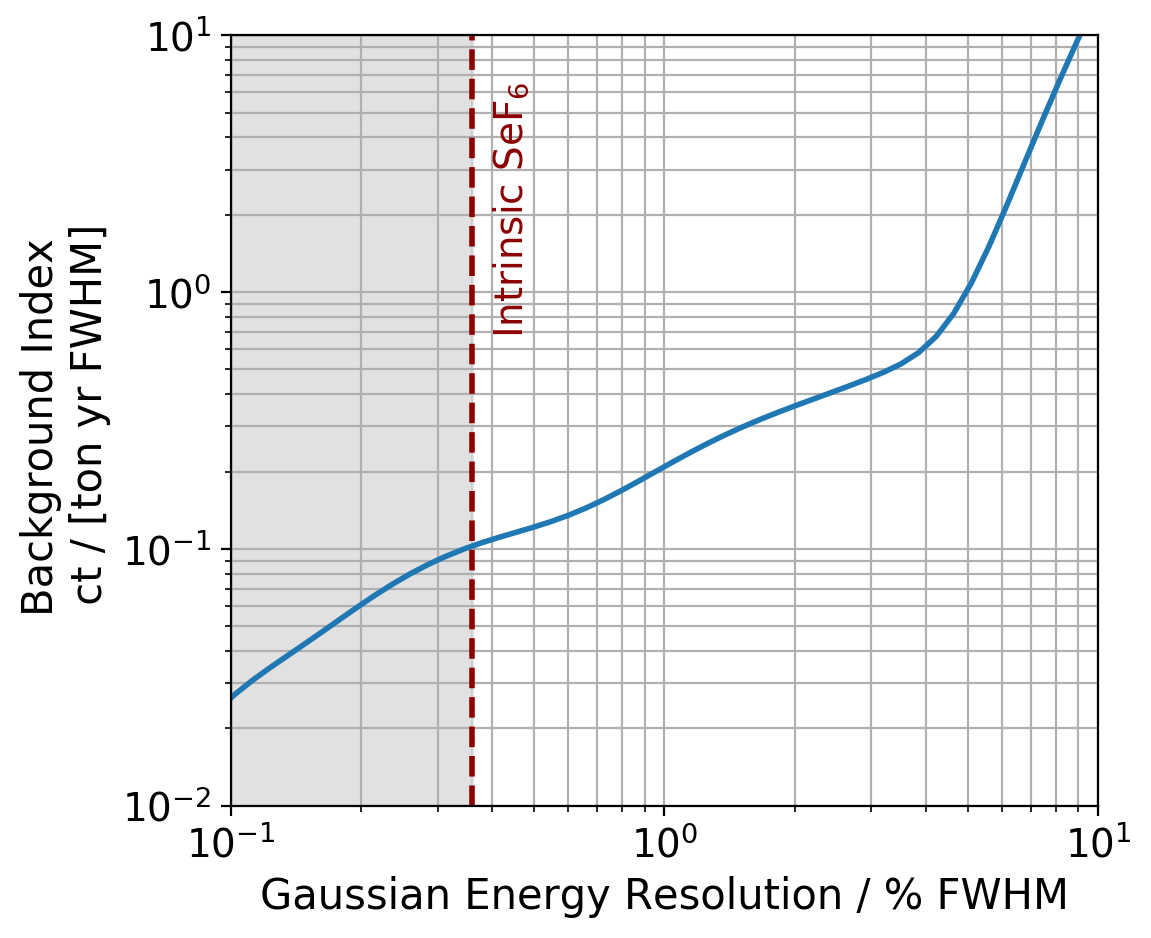} 

\caption{Left: Background index ratio around the $^{82}$Se Q-value compared to around the $^{136}$Xe Q-value as a function of Gaussian energy resolution. Right: Approximate absolute background index for a 100~kg-scale SeF$_6$ TPC, obtained using the rate relative to xenon and the background index projections from  \cite{Martin-Albo:2015rhw} \label{fig:BGIndex}}
\end{center}
\end{figure}

By convolving the total gamma rate with a finite Gaussian around the Q-value representing the achievable energy resolution, we can compare the expected background rate as a function of energy resolution in similarly precise detectors.  We see that, if similar resolution can be achieved with ion counting as has been demonstrated using  electroluminescence, the expected background index in a 100~kg SeF$_6$ detector may be between 10 and 25 times smaller than in a HPGXe. This is shown in Fig.~\ref{fig:BGIndex}, left.  Normalizing to the predictions of \cite{Martin-Albo:2015rhw}, we can estimate the absolute background index of such an experiment, shown in Fig. \ref{fig:BGIndex}, right. We find that background indices of around 0.2 ct/[ton year FWHM] can be achieved, if the demonstrated energy resolutions from xenon gas can be replicated.  Even relatively poor energy resolution of around 2\% FWHM would yield background index significant better than 1 ct/[ton year FWHM], which would be world-leading within the field of neutrinoless double beta decay.

These numbers, though highly preliminary, suggest that the improvements implied by conducting a gas-phase search using an isotope with Q-value above the most common bismuth and thallium lines are substantial.  Such an experiment would be almost background free at the ton-scale, and may achieve background indices unrivaled by any contemporary running or proposed experiment.  It is also notable that our estimates here related to a 100~kg detector. A ton-scale device would be expected to exhibit an even better background index, through improved surface-to-volume ratio.

\section{Sensitivity Estimates}

Using the background estimates of Section \ref{sec:Backgrounds} we can estimate the sensitivity to neutrinoless double beta decay half life T$_{1/2}$.   We express this sensitivity using two widely used conventions, to facilitate comparison with projections from other experiments:
\begin{itemize}
    \item {\bf Median sensitivity at 90\% CL.} This is defined as the median limit that would be set at 90\% CL given experiments where there is no true signal.
    \item {\bf 3$\sigma$ discovery potential.} This is defined as the true signal size which would be discoverable at 3$\sigma$ in 50\% of trials of a given experiment.
\end{itemize}
Although both are presented, if one considers the search for neutrinoless double beta decay to be a quest for discovery rather than a limit setting exercise, the latter should be emphasized in comparisons of experimental techniques.  

To evaluate these sensitivities we follow the heuristic counting approach described in \cite{Agostini:2017jim}, considering events within an optimized region of interest around the Q-value. Our projections are likely rather conservative, with additional sensitivity achievable using a full likelihood fit implemented as a function of event energy and other signal-vs-background discriminants. We also use background estimates derived for a 100~kg device, though in practice a larger experiment will naturally exhibit lower background due to beneficial scaling effects.

The median sensitivity and discovery potential are directly determined by the expected background rate, which in turn depends on the achievable energy resolution. We consider three cases of interest: 1) the intrinsic resolution of SeF$_6$ as estimated in previous sections, which is $\sim$0.4\% FWHM; 2) the resolution demonstrated in high pressure xenon gas with NEXT-NEW, which is $\sim$0.7\% FWHM; 3) a pessimistic energy resolution of 2\% FWHM, assumed to be dominated by instrumental effects.

\begin{figure}[t]
\begin{center}
\label{Lego}

\includegraphics[width=0.68\columnwidth]{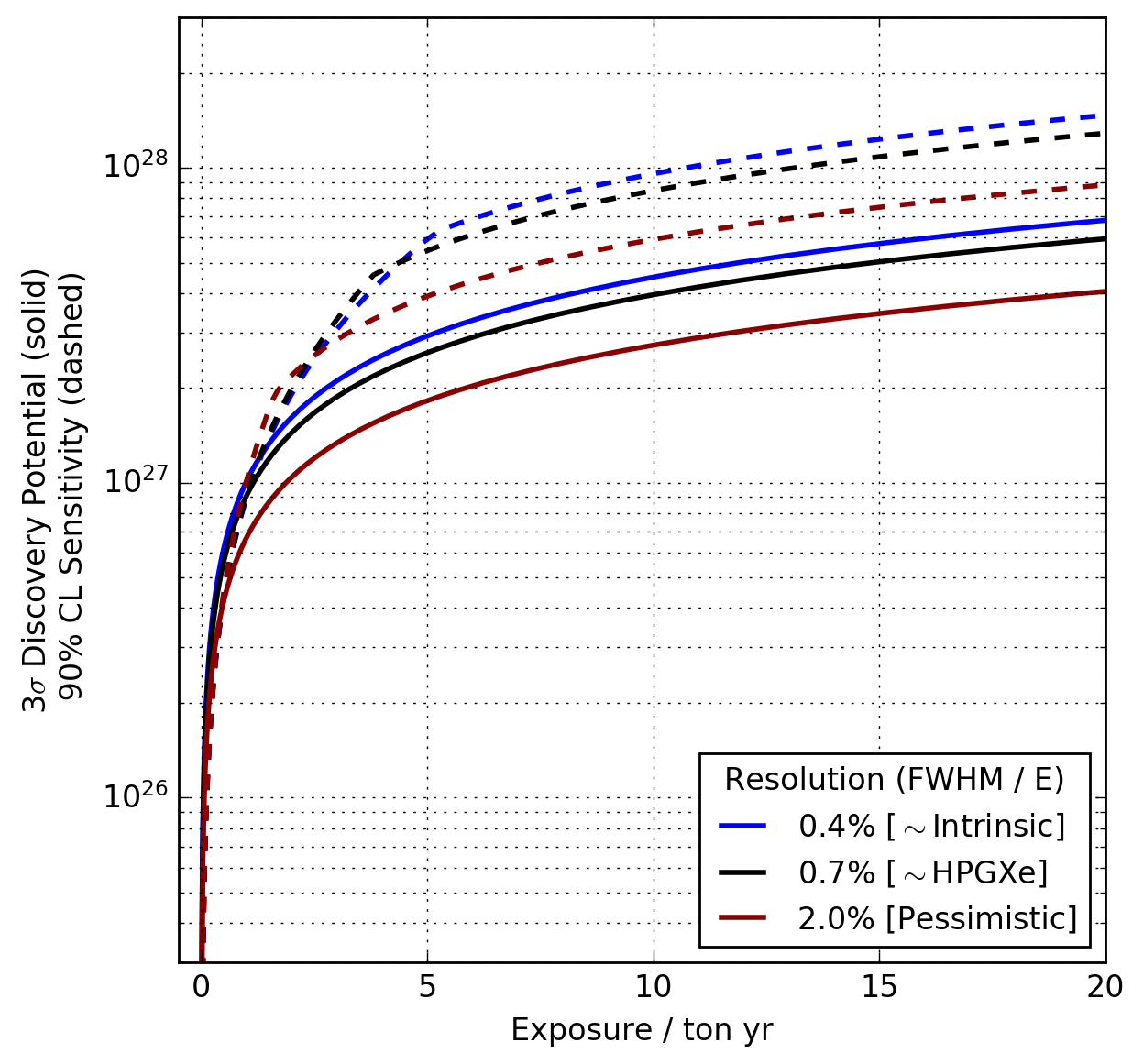} 

\caption{Estimated sensitivity and discovery potential of an enriched SeF$_6$ TPC using various assumptions about the achievable energy resolution.  \label{fig:sensitivity}}
\end{center}
\end{figure}

Figure~\ref{fig:sensitivity} shows both discovery potential and sensitivity on a logarithmic scale.  After 10 ton-years of exposure, 90\% CL sensitivities of around 10$^{28}$ yr are achievable in the most optimistic cases. These correspond to 3$\sigma$ discovery potentials of 5$\times$10$^{27}$ years.  This exceeds the expected sensitivity in xenon-based TPC  experiments at similar exposure, which is particularly notable since a given mass of enriched SeF$_6$ contains only 136/196 as many decay candidates of a similar mass of enriched xenon.  This sensitivity is achieved primarily through the intrinsically low background levels near the high Q-value of selenium, combined with the characteristic gas-phase benefits of excellent energy resolution and topological discrimination.   Even in the more conservative scenario where 2\% is the maximum achievable energy resolution, discovery potentials significantly in excess of $10^{27}$ years are expected with 10 ton years of exposure.  

\section{Conclusions}

We have presented a concept for a novel neutrinoless double beta decay experiment based on enriched selenium hexafluoride gas.  Such a device would operate as an ion TPC and, through the high Q-value of selenium, may achieve lower background indices than are achievable with other types of neutrinoless double beta decay TPC based on liquid and gaseous xenon.  

By comparison with simulations of SF$_6$ we estimate that expected intrinsic energy resolution is likely to be substantially better than 1\% at $Q_{\beta\beta}$ with charges distributed across various species of positive and negative ions. Individual counting of these ions may further improve the energy resolution beyond what that suggested by the conventionally defined Fano factor, though not by a substantial factor.  The micro-physics of ion formation and transport in this system is highly non-trivial, and further work, both computational and experimental, is required.

Performance of topological algorithms for identification of double beta decay candidates in SeF$_6$ gas appears improved relative to the equivalent algorithms applied in high pressure xenon gas. This owes primarily to the higher Q-value of selenium and the reduced multiple scattering in a lower-Z material.

Radioactive background rates estimated based on simulations of a structure similar to the NEXT-100 xenon gas TPC suggest that the background index of a SeF$_6$ TPC would be in the range 0.1<b<0.5 in units of counts per ton per year per FWHM, with a relatively weak dependence on energy resolution.  Assuming energy resolution better than around 2\% FWHM can be achieved to reject events from two-neutrino double beta decay, an extremely low background neutrinoless double beta decay TPC appears achievable.  

With such a detector, discovery potentials in the range of 5$\times$10$^{27}$ years may be achievable with 10 ton-years of exposure. This is somewhat beyond the capability of proposed xenon gas experiments and far beyond the capability of proposed liquid xenon experiments operating with similar ton-year exposures.

Much remains to be explored and understood about the SeF$_6$ ion TPC concept, and many of our speculations and estimates here demand experimental verification.  However, our preliminary estimates show that this may be a highly promising avenue for R\&D toward ultra-low background, ton-scale neutrinoless double beta decay detectors.

\acknowledgments
We thank Diego Gonzalez Diaz and Stephen Biagi for invaluable discussions regarding SF$_6$ and SeF$_6$ micro-physics and simulation, and Nu Xu for discussions about prospects for SeF$_6$ experiments.  We also thank the NEXT collaboration for their input, and for allowing our use of NEXT simulation techniques within the context of this work.  J.~Renner acknowledges the support of a Juan de la Cierva postdoctoral contract from the Spanish Ministry (MINECO), reference FJCI-2015-25861, and the program Subvenciones a la Excelencia Cient\'{i}fica de Juniors Investigadores of the Generalitat Valenciana under grant number SEJI/2017/011.  F.~Psihas is supported by a ConTex postdoctoral fellowship from the UT System and Conacyt.  N.~L\'opez-March received funding from the European Union's Framework Programme for Research and Innovation Horizon 2020 (2014-2020) under the Marie Sk{\l}odowska-Curie Grant Agreement No.\ 740055.  Y.~Mei was supported by Laboratory Directed Research and Development (LDRD) funding from Berkeley Lab, provided by the Director, Office of Science, of the U.S.\ Department of Energy under Contract No.\ DE-AC02-05CH11231.  The Rare Events Searches and Techniques group at the University of Texas at Arlington receives support from the Department of Energy under grant DE-SC0017721.
\bibliographystyle{JHEP}
\bibliography{main}

\end{document}